\documentclass{aa}  

\usepackage{dcolumn}
\newcolumntype{d}[1]{D{.}{.}{#1}}  
\usepackage{graphicx}
\usepackage{xcolor,colortbl}
\usepackage{txfonts}
\usepackage{natbib}
\usepackage[colorlinks=true,urlcolor=blue,citecolor=blue,linkcolor=blue,breaklinks=true]{hyperref}
\begin{document}

   \title{The extremely strong non-neutralized electric currents of the unique solar active region NOAA\,13664}

   \author{I. Kontogiannis
          \inst{1}
          }

   \institute{Leibniz-Institut f\"ur Astrophysik Potsdam (AIP) Germany, An der Sternwarte 16 14482 Potsdam Germany \\
              \email{ikontogiannis@aip.de}
             }

   \date{26 September 2024}

 
  \abstract
   {In May 2024, the extremely complex active region National Oceanic and Atmospheric Administration (NOAA)\,13664 produced the strongest geomagnetic storm since 2003.}
   {The aim of this study is to explore the development of the extreme magnetic complexity of NOAA\,13664 in terms of its photospheric electric current.}
   {The non-neutralized electric current is derived from photospheric vector magnetograms, provided by the Helioseismic and Magnetic Imaged onboard the Solar Dynamics Observatory. The calculation method is based on image processing, thresholding and error analysis. The spatial and temporal evolution of the non-neutralized electric current of the region as well as its constituent sub-regions is examined. For context, a comparison with other complex, flare-prolific active regions is provided.}
   {Active region NOAA\,13664 was formed by the emergence and interaction of three sub-regions, two of which were of notable individual complexity. It consisted of numerous persistent, current-carrying magnetic partitions that exhibited periods of conspicuous motions and strongly increasing electric current at many locations within the region. These periods were followed by intense and repeated flaring. The total unsigned non-neutralized electric currents and average injection rates reached $5.95\cdot10^{13}$\,A and $1.5\cdot10^{13}$\,A/day, and were the strongest observed so far, significantly surpassing other super-active regions of Solar Cycle 24 and 25.}
   {Active region NOAA\,13664 presents a unique case of complexity. Further scrutiny of the spatial and temporal variation of the net electric currents during the emergence and development of super-active regions is paramount to understand the origin of complex regions and adverse space weather.}


\titlerunning{Non-neutralized electric currents of NOAA\,13664}
\authorrunning{Kontogiannis I.}
\maketitle
%

\section{Introduction}

Between 10 and 13~May 2024, the strongest geomagnetic storm since 2003 took place. The recorded geomagnetic index reached Kp = 9 and Dst = -412\,nT at 03:00 UTC on 11 May\footnote{\url{https://www.swpc.noaa.gov/}} and strong aurora was observed, even from low geographic latitudes. This activity was the result of intense and repeated eruptions in the Sun. Starting on 8 May 2024 and within six days, the source region, National Oceanic and Atmospheric Administration (NOAA) 13664, produced 8 X-class flares and several M- and C-class flares, many of which were associated with coronal mass ejections (CMEs) making it the most flare-prolific active region in Solar Cycle 25 and one of the most eruptive ones ever recorded. 

Active regions are strong, extended and often highly complex magnetic structures, whose formation is the result of the emergence of magnetic flux from the solar interior \citep{2015LRSP...12....1V}. Initially appearing as bipolar features, their peak size and complexity vary and only some of them evolve into highly eruptive ones. These regions consist of closely packed opposite magnetic polarities, giving rise to $\delta$-spots \citep{1965AN....288..177K} and strong polarity inversion lines (PILs) with intense shearing motions \citep[see e.g.,][]{2007ApJ...655L.117S,2023AdSpR..71.2017K}. Their formation could be the result of kink instability in a largely bipolar region or follow interactions of smaller flux systems, which could be part of the same or of different sub-photospheric structures \citep[see e.g.,][and references therein]{kniznik18,levens23}. The aforementioned studies attest to the importance of the photospheric manifestations of flux emergence as possible indicators of the sub-photospheric origin of complex regions. The same manifestations of the various types of magnetic interactions inferred from photospheric magnetograms are also considered drivers of flares and CMEs \cite{toriumi17}

   \begin{figure*}
   \centering
   \includegraphics[width=19cm]{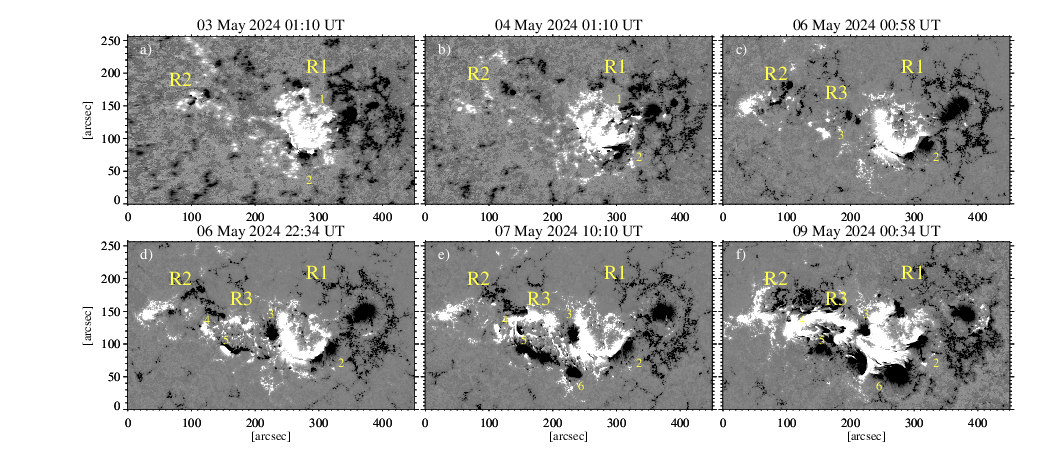}
      \caption{ Photospheric magnetograms at various instances of the evolution of active region NOAA\,13664. The three sub-regions are indicated with R1,R2, and R3, while numbers from ``1'' to ``6'' indicate features of interest (see text).}
         \label{fig:magevo}
   \end{figure*}

The resulting complex magnetic configurations require the presence of strong electric currents, which can be routinely calculated from vector photospheric magnetograms \citep{Leka96}. In fact, observations and numerical simulations show that photospheric electric currents can be neutralized only in the case of isolated symmetric sunspots, but in case of emergence of magnetic flux and close proximity of opposite magnetic polarities this symmetry breaks and strong non-neutralized electric currents develop \citep{geo_titov_mikic12,torok14,dalmasse15}. Since most regions deviate from a simple potential (current-free) state, some amount of net electric currents is almost always present. However, scrutiny of large samples of active regions \citep{kontogiannis17,2024ApJ...970..162K} has shown that for regions with no PILs and no signs of flux emergence, most of this electric current is neutralized. On the contrary, the latter is significantly high when PILs and strong shear start to develop. Thus, $\delta$-regions exhibit significantly higher non-neutralized electric current, as well as build-up rates. Additionally, the amount of these electric currents is strongly correlated with flare output and, in the case of eruptions, with CME speed and acceleration \citep{kontogiannis17,2017ApJ...846L...6L,2019MNRAS.486.4936V,kontogiannis19,2020ApJ...893..123A,2024ApJ...961..148L}.

Although several mechanisms have been put forward to explain the initiation of strong flares and CMEs \citep{2018SSRv..214...46G}, it is firmly established that the source regions of major eruptions present considerable complexity. The strong association between magnetic complexity, electric currents and capacity to generate adverse space weather effects necessitate the analysis of how non-potentiality builds up in super-active regions. For this reason, several works have focused on the emergence and evolution of outstanding active regions such as NOAA\,11158, 11429, 12192, and 12673 from varying standpoints \citep[see e.g.,][]{tziotziou13,2015Sun,2016ApJ...817...14P,2018A&A...612A.101V,2019ApJ...871...67C}. The present study focuses on the evolution of the non-neutralized electric currents of the super-active region NOAA\,13664 and its constituents. The measured electric currents are the strongest ones measured since 2010.

\section{Data and Analysis}

The Helioseismic and Magnetic Imager \citep[HMI;][]{hmischerrer,hmischou} onboard the Solar Dynamics Observatory \citep[SDO;][]{sdo} provides photospheric vector magnetograms, derived from spectropolarimetric observations at the Fe\,I\,6173\,\AA\ spectral line. For this study the Cylindrical Equal Area (CEA) version of the the Space Weather HMI Active Region Patches data \citep[SHARP;][]{bobra14} definitive data is used. This product provides the $B_{r}$, $B_{p}$ and $B_{t}$ (equivalent to the three Cartesian components $B_{z}$, $B_{x}$ and $B_{y}$), deprojected and remapped to the solar disk center.

The method used to derive the non-neutralized electric current was developed by \citet{geo_titov_mikic12} and further applied on sizable active region samples in \cite{kontogiannis17} and \citet{2024ApJ...970..162K}. The map of the vertical component of the magnetic field, $B_{r}$, is subdivided into unipolar partitions using thresholds for $B_{r}$ (100\,G), the minimum partition size (5.3\,Mm$^2$), and the minimum magnetic flux per partition (5$\times10^{19}$\,Mx). The total vertical electric current within each partition and the propagated error are calculated using the differential form of Ampere's law. In order to provide a baseline for the numerical effects and errors introduced in the calculation of the electric current, the ``electric current for the potential magnetic field'' $I_{pot}$ is derived, as follows. Using the $B_{r}$ component, the potential (current-free) magnetic field at the photosphere is calculated using the extrapolation method by \citet{alissandrakis81}. Then, by means of the Ampere's law, the corresponding electric current is derived, which in the general case deviates from zero due to numerical effects. Magnetic partitions that carry non-neutralized electric current satisfy the conditions $I>5 \times I_{pot}$ and $I>3 \times \delta I$. Thus, for each SHARP cut-out (such as the ones shown in Fig.~\ref{fig:magevo}) the electric-current neutralized partitions are eliminated, leaving only the non-neutralized magnetic partitions, (if any), each of them carrying net electric current $I_{NN}^{i}$ and magnetic flux $\Phi_{NN}^{i}$. From the distribution of $I_{NN}^{i}$, the total unsigned non-neutralized electric current $I_{NN,tot}$ = $\Sigma$$|I_{NN}^i|$ and the total unsigned magnetic flux $\Phi_{NN,tot}$ = $\Sigma$$|\Phi_{NN}^i|$ of the current-carrying part of the region are determined \citep{kontogiannis17,2024ApJ...970..162K}. The centroids of the non-neutralized partitions were used to track the evolution of the most persistent of them.


   \begin{figure*}
   \centering
   \includegraphics[width=18cm]{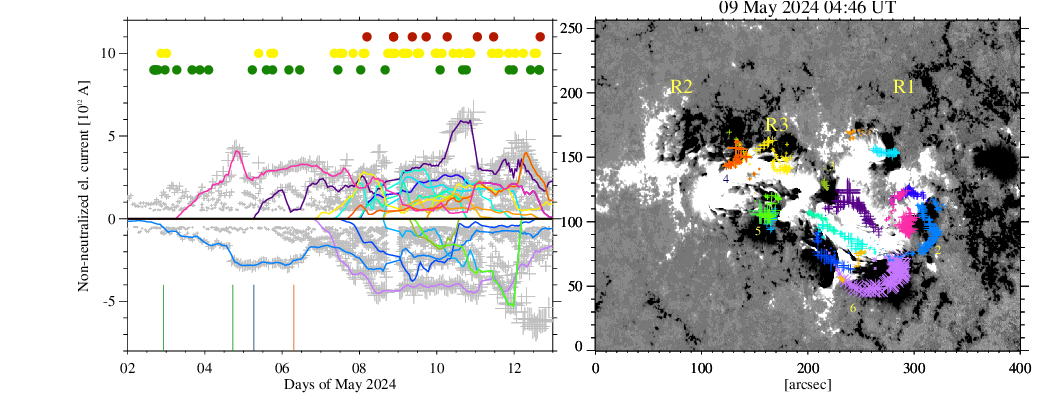}
      \caption{Left: The electric currents $I_{NN}$ carried by the non-neutralized partitions of the active region, as a function of time (gray crosses). The partitions that could be tracked over large parts of the time series are connected with colored solid lines. Only partitions with $I_{NN}$ that exceeds $0.5\cdot10^{12}$\,A are plotted, for clarity. Flaring activity is indicated as in Fig.~\ref{fig:nntot}. Vertical lines indicate the two emergence events of R2 (green), the emergence of R3 (blue) and the time of collision onset between the negative polarity of R3 with the positive of R1 (red). Right:Trajectories of the non-neutralized partitions plotted in the left panel, overplotted on a photospheric magnetogram of NOAA\,13664. Different colors indicate correspondence with the curves in the left panel while larger ``+'' signs indicate stronger absolute value of $I_{NN}$. The locations included in Fig.~\ref{fig:magevo} are also provided here for reference.}
         \label{fig:nn_partitions}
   \end{figure*}
   
\section{Results}

Fig.~\ref{fig:magevo} contains snapshots from the evolution of NOAA\,13664. The region was already evolved when it rotated into view, on 30 April 2024. It received a NOAA identifier on 2 May, when it was sufficiently away from the western limb. Initially (Fig.~\ref{fig:magevo}(a)), it consisted of a well-developed bipole (R1), which was still exhibiting emergence (``1''). The negative polarity of R1 developed into a well-formed sunspot and the positive polarity consisted of several smaller ones. In addition to the main polarities of R1 there existed also a parasitic bipole (``2''), which was emerging adjacent to the positive magnetic polarity, exhibiting strong shearing motions. This part kept evolving throughout the entire observation span (see also panels (b)-(f), ``2''). In addition to the evolving R1, another, weaker sub-region, R2, was emerging at the north-western part of the FOV, which exhibited a stronger emergence event on 04 May 2024. 
   
Two days later, on 6 May 2024 (Fig.~\ref{fig:magevo}(c)), a third sub-region, R3, started to emerge in between R1 and R2, which would shape drastically the evolution of NOAA\,13664. After the initial emergence of R3, and as it started expanding, the first negative polarity, ``3'', of the sub-region moved towards the positive polarity of R1. Six hours later, the bulk of R3 started to emerge. While the initial emergence of R3 was in a Hale orientation (just like the R1 and R2 sub-regions), the two new polarities ``4'' and ``5'' were anti-Hale (Fig.~\ref{fig:magevo}(c) and (d)). Apart from the separation motion, the elongated polarities themselves exhibited also apparent rotation/twisting, indicating the emergence of considerably twisted magnetic flux \citep{2003A&A...397..305L,2021NatCo..12.6621M}. The emergence of R3 pushed the pre-existing negative footpoint ``3'' further into the positive polarity of R1. The negative polarity ``6'' of R3 also directed toward R1, moving in parallel to its positive polarity, while the positive polarity, ``4'', of R3 collided with the negative polarity of R2 (Fig.~\ref{fig:magevo}(e)). The result of this interaction was the complex configuration seen in Fig.~\ref{fig:magevo}(f), with multiple closely neighboring, highly deformed magnetic polarities, with strong shearing motions extending over most of the active region area. As a result, repeated flaring started on 07 May, with several frequent X-class flares after 8 May, including an X5.8 and X8.7 on 11 and 14 May correspondingly, the latter being the strongest up to that point flare of Solar Cycle 25 on the Earth-facing side of the Sun. The CMEs associated with the X-class flares that took place until 11 May were geoeffective and caused the strongest geomagnetic events since 2003.

The spatial and temporal variability of the non-neutralized electric current is shown in more detail in  Fig.~\ref{fig:nn_partitions}. Overall the region comprised many persistent and strongly non-neutralized partitions. Both their number and their associated $I_{NN}$ increased, as indicated by the increasing density in the left panel of Fig.~\ref{fig:nn_partitions}, following the emergence events of R2 (vertical green lines) and, mainly, the emergence of R3 (vertical blue line). This effect was more dramatic when R3 started to approach and merge with R1 (vertical orange line). 

The right panel of Fig.~\ref{fig:nn_partitions} shows the trajectories of the centroids of the most persistent and strongly non-neutralized partitions. These are found in regions of intense interaction and electric current injection (see also Fig.~\ref{fig:magevo}). Some of them stand out in terms of increasing $I_{NN}$ and displacement, such as the ones at ``2'' (light blue and magenta), ``3'' (green) and ``6'' (lilac), as well as at ``4'' (orange) and ``5'' (light green). As a result of this dynamic evolution, the location of the region that hosted the strongest electric current was changing with time. Initially it was located at the PIL within R1 (``2''), on 8 May it was at the PIL between R1 and R3 (``3'' and ``6'') and after 11 May it shifted to the interaction region between R3 and R2 (''4`` and ``5''). Periods of increasing $I_{NN}$ at these locations seem to have led to intense flaring, such as the repeated X-class flares after 8 May and the X5.8 flare on 11 May (Fig.~\ref{fig:nn_partitions}, left panel). The inferred association between local development of non-neutralized current and flaring is worth exploring in more detail in the future by applying this methodology to a larger sample of active regions.

   \begin{figure}
   \centering
   \includegraphics[width=9cm]{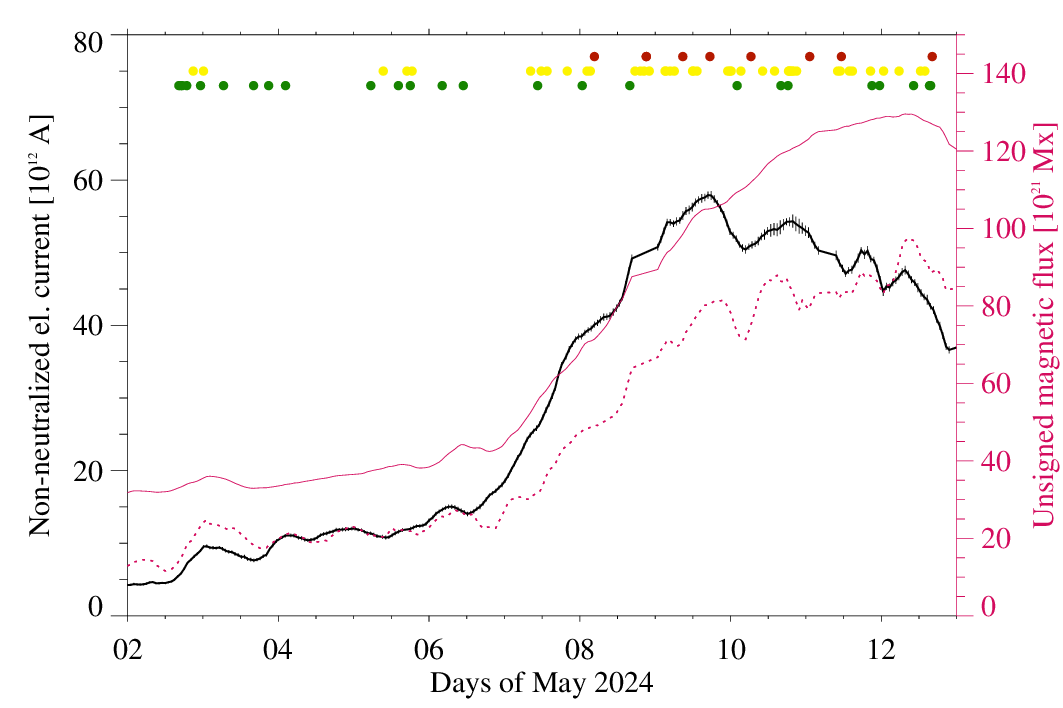}
      \caption{Time series of the total unsigned non-neutralized electric currents, $I_{NN,tot}$ (black). Overplotted are also the total unsigned magnetic flux $\Phi$ (solid pink) and the magnetic flux of the non-neutralized partitions $\Phi_{NN,tot}$ (dotted pink). Green, yellow and red dots indicate occurrences of C-, M- and X-class flares, provided by NOAA SWPC. The time axis is in days, starting on 2 May 20224.}
         \label{fig:nntot}
   \end{figure}

The temporal evolution of the total unsigned non-neutralized electric current and the magnetic flux is shown in Fig.~\ref{fig:nntot}. The region already possessed very strong, on the order $5\cdot10^{12}$\,A, net currents, when it rotated into view. These net currents were already increasing up to 6 May 2024, predominantly due to the evolution of R1 and the shearing motions of ``2'', before the emergence of R3 and the merging of all three regions (see also Fig.~\ref{fig:nn_partitions}). Up to that point, the average increase rate of $I_{NN,tot}$ for the entire region was $2.16\cdot10^{12}$\,A/day and $I_{NN,tot}$ had already exceeded $10\cdot10^{12}$\,A. The observed increase rate of the region (essentially still the sub-region R1) was already at the upper range of values for $\delta$-regions, as determined by \citep{2024ApJ...970..162K}, and in terms of net electric current the region already had the potential to produce an X-class flare \citep{kontogiannis17}. In fact it produced at least 5 M- and 15 C-class flares before the sharp increase in complexity that took place after 6 May 2008.

After 6 May 2024, a dramatic increase in the magnetic flux, as well as in $I_{NN,tot}$ and $\Phi_{NN,tot}$ was observed. The total unsigned non-neutralized electric current peaked on 09 May 2024 at 17:46\,UT, the rate during the fast increase was $1.5\cdot10^{13}$\,A/day and the peak value was $5.95\cdot10^{13}$\,A. To the author's knowledge, both the increase rate and the peak $I_{NN,tot}$ value exceed measurements for any other previous active region of the last two solar cycles. \citet{2024ApJ...970..162K} quantified the current-carrying part of active regions using the ratio between the non-neutralized and the total unsigned magnetic fluxes. For NOAA\,13664 this ratio reached up to 0.80, which exceeds the 75th percentile of the values calculated in \citet{2024ApJ...970..162K}, indicating that most of the magnetic flux of the region was current-carrying.

At this point, a comparison of the total unsigned non-neutralized electric current developed in NOAA\,13664 with that of other regions will help put the former into the context of other similarly eruptive active regions. The two well-studied regions NOAA\,11158 and NOAA\,12673 produced the first and the strongest X-class flares of Solar Cycle 24, respectively \citep[see][for informative reviews of the literature on these two regions]{tziotziou13,2024A&A...682A..46P}. In both regions the interaction between different flux systems, which eventually led to complex configurations, could be monitored from the beginning. For NOAA\,11158 these two flux systems emerged with a very small time lag and then merged into a quadrupolar configuration with a well-formed and highly-sheared PIL \citep{toriumi17}. Active region NOAA\,12673, on the other hand, consisted of a well-formed sunspot, which was already in its decay phase. The emergence of highly deformed magnetic flux and its collision with the pre-existing sunspot created the complex configuration of NOAA\,12673 \citep{2018A&A...612A.101V}. 

\begin{figure}
   \centering
   \includegraphics[width=9cm]{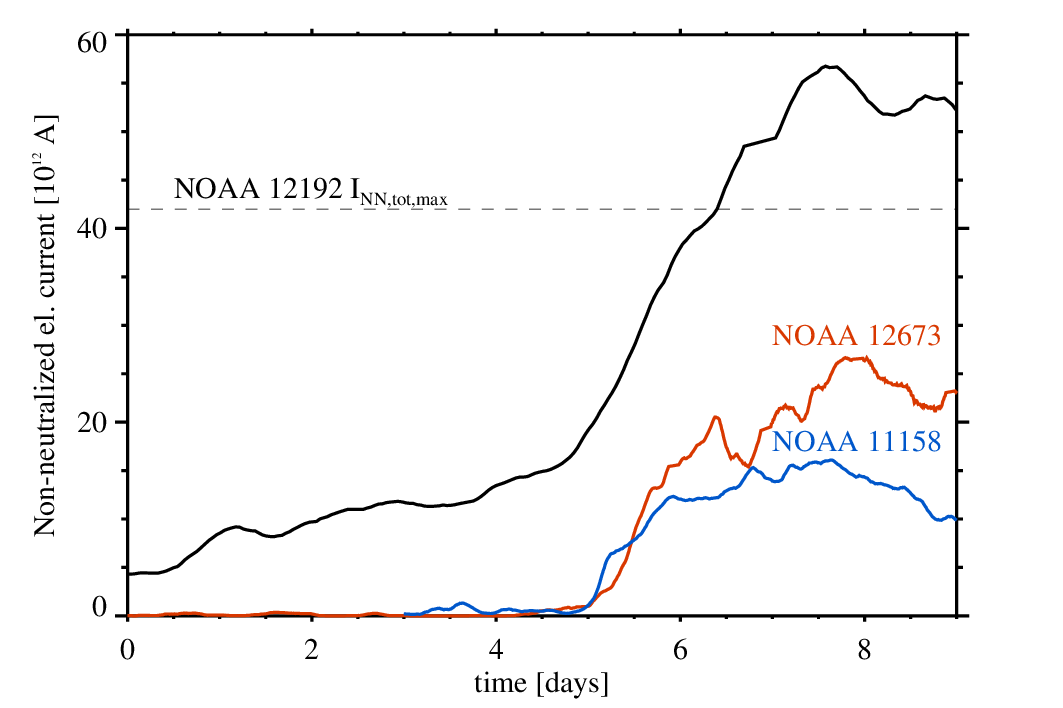}
      \caption{Time series of $I_{NN,tot}$ of NOAA\,11158 (blue) and NOAA\,12673 (red) in comparison to NOAA\,13664 (black). To facilitate comparison, the time series were shifted in time and the and the maximum $I_{NN,tot}$ reached by NOAA\,12192 is marked with the dashed, horizontal line.}
         \label{fig:nn_regions}
   \end{figure}

Although the two regions were also flare prolific and eruptive, the difference with NOAA\,13664 in terms of $I_{NN,tot}$ is clear (Fig.~\ref{fig:nn_regions}). For NOAA\,11158, during the emergence of the first two bipoles $I_{NN,tot}$ reached  $10^{12}$\,A and further evolution of the region led to the increase of $I_{NN,tot}$ up to $16\cdot10^{12}$\,A. For NOAA\,12673, the pre-existing sunspot was carrying very weak, if at all, non-neutralized electric current and only after the emergence events of the following days did $I_{NN,tot}$ start increasing dramatically. Both NOAA\,12673 and NOAA\,13664 exhibited very fast magnetic flux emergence \citep[][see also\footnote{\url{http://hmi.stanford.edu/hminuggets/?p=4216}}]{2017RNAAS...1...24S} and both resulted from the interaction between consecutive emergence events in the vicinity of already established regions. However, NOAA\,12673 consisted initially of an isolated sunspot, while NOAA\,13664 was already a highly complex $\delta$-region. Before the interaction and merging with R3 and R2, the pre-existing sub-region R1 was already, in terms of electric current and activity potential, at a level comparable to that of a fully-developed region like NOAA\,11158 (despite their different morphology). 

At its full extent, NOAA\,13644 became the second most extended region of the past two decades, second only to NOAA\,12192 \citep{2015Sun}. The latter also contained very strong electric current (see dashed line in Fig.~\ref{fig:nn_regions}), but although NOAA\,13664 reached the 87\% of the area of NOAA\,12192 it contained 28\% strongest non-neutralized electric current.

\section{Conclusions and Discussion}

The source region of the May 2024 eruptions, NOAA\,13664, exhibited extremely strong non-neutralized electric currents, likely the strongest ever measured since the start of SDO observations. These electric currents exhibited strong spatial and temporal variability, with very high injection rates, as a result of a combination of flux emergence and interactions between different, complex flux systems. Repeated and strong flaring followed periods of intense increase of the electric current, carried by the persistent and strongly non-neutralized partitions of the region.       

Although the formation of highly complex regions is attributed to various forms of flux systems interactions \citep[see e.g.,][]{toriumi17,2018A&A...612A.101V,2019ApJ...871...67C}, two features seem to differentiate NOAA\,13664, making it a truly unique case. First, the already established magnetic environment exhibited strong non-potentiality. When it rotated into view, NOAA\,13664 was already a highly non-potential active region, developing already strong non-neutralized electric current, well before the flux emergence that followed. Second, the emergence and development of two sub-regions, including an anti-Hale one, injected significant electric currents. Based on the established statistical relationship of flaring activity and non-neutralized electric currents \citep{kontogiannis17,2024ApJ...970..162K} it is likely that both the pre-existing region and the anti-Hale sub-region would produce strong flares even if they were isolated ones. The resulting configuration was the second most extended active region since 2010, following active region NOAA\,12192. However, the severity of electric current injection in the former was such, that it significantly surpassed the latter in terms of non-neutralized electric current.

The formation of super-active regions and $\delta$-regions in general is an open subject, relevant to fundamental mechanisms with which the magnetic field emerges and interacts with the turbulent plasma \citep{2019LRSP...16....3T}. Additionally, these regions are extremely interesting in terms of induced space weather, not only for producing very strong eruptive phenomena in their prime, but also for continuing to impact solar magnetism and shaping the space environment during their long decay \citep[see e.g.,][]{2010ApJ...714.1672K,2017ApJ...840..100M,2020ApJ...904...62W}. 
The methodology used in this study combines summary information of the active region complexity, in terms of the total unsigned non-neutralized electric current, with the local spatial and temporal evolution of the non-neutralized current carried by sub-regions and their constituents. A future step would be to apply this novel approach to larger samples of regions to explore why some of them evolve into super-active ones, how they decay and whether there exist features that would improve our ability to predict flares and CMEs. Regarding NOAA\,13664, only the later stages of emergence were captured, and the region has recurred already twice, producing considerable activity during its decay. Therefore, magnetic field observations from more than one vantage points are necessary to track longer the extended activity complexes and capture crucial parts of their emergence, evolution and decay  \citep{2023A&A...673A..31S,2024A&A...685A..28M}.

\begin{acknowledgements}
     I would like to thank the anonymous referee for providing remarks that motivated a clearer presentation of the results. This work was supported by the \emph{Deut\-sche For\-schungs\-ge\-mein\-schaft (DFG)\/} project number KO 6283/2-1. Data from SDO/HMI are courtesy of NASA/SDO and the AIA, EVE, and HMI science teams and are publicly available through the Joint Science Operations Center at the \url{jsoc.stanford.edu}.)
\end{acknowledgements}

%
%

\bibliography{references}{}
\bibliographystyle{aa}

\end{document}